\documentclass{aa}
\usepackage{soul}
\usepackage{graphicx}
\usepackage{txfonts}
\usepackage{hyperref}
\usepackage{natbib}
\usepackage{makecell}
\usepackage{xcolor}
\usepackage{makecell}
\usepackage{lipsum}
\usepackage{float}
\usepackage{needspace}
\usepackage{pdflscape}
\usepackage{placeins}
\usepackage{booktabs} 
\usepackage[normalem]{ulem}
\renewcommand{\arraystretch}{1.5}

\newcommand{\XP}{xallarap}
\newcommand{\PX}{parallax}
 
\newcommand{\FS}{FS}
\newcommand{\LD}{LD}
\newcommand{\sS}{Planet-S2}
\newcommand{\sP}{Planet-S1}
\newcommand{\ET}{ET}
\newcommand{\PriorSize}{Prior B}
\newcommand{\PriorFlux}{Prior A}
\newcommand{\Prior}{\mathrm{Prior}}

\newcommand{\affOAUW}{Astronomical Observatory, University of Warsaw, Al. Ujazdowskie 4, 00-478 Warszawa, Poland}
\newcommand{\affWARWICK}{Department of Physics, University of Warwick, Coventry CV4 7AL, UK}

\newcommand{\affVillanova}{Villanova University, Department of Astrophysics and Planetary Sciences, 800 Lancaster Ave., Villanova, PA 19085, USA}

\begin{document} 

\title{Incorporating physical source parameters into microlensing modeling}
   
\author{Mateusz~J.~Mr{\'o}z\inst{\ref{OAUW}}\and Radosław~Poleski\inst{\ref{OAUW}}\and Andrzej~Udalski\inst{\ref{OAUW}}\and 
Jan~Skowron\inst{\ref{OAUW}}\and
Paweł~Pietrukowicz \inst{\ref{OAUW}}\and
Michał~K.~Szyma\'nski \inst{\ref{OAUW}}\and
Przemek~Mróz\inst{\ref{OAUW}}\and
Mariusz~Gromadzki\inst{\ref{OAUW}}\and
Patryk~Iwanek\inst{\ref{OAUW}}\and
Szymon~Koz\l{}owski\inst{\ref{OAUW}}\and
Milena~Ratajczak\inst{\ref{OAUW}}\and
Krzysztof~A.~Rybicki\inst{\ref{OAUW}}\and
Dorota~M.~Skowron \inst{\ref{OAUW}}\and
Igor~Soszy\'nski \inst{\ref{OAUW}}\and
Krzysztof~Ulaczyk \inst{\ref{WARWICK}}\and
Marcin~Wrona \inst{\ref{Villanova}, \ref{OAUW}}\and
Zofia~Buzik \inst{\ref{OAUW}}
}

\institute{ \affOAUW\label{OAUW} \and
\affWARWICK \label{WARWICK} \and
\affVillanova \label{Villanova}}

\abstract{Modeling of complex microlensing events suffers from many difficult-to-disentangle degeneracies. This is especially the case for orbital motion of the source in a binary system, the so-called xallarap effect. To address the degeneracies inherent in xallarap modeling, we developed a novel approach that directly samples the physical parameters of the source stars (initial mass, evolutionary phase, metallicity, distance, and reddening) during MCMC fitting. In our approach the physical parameters of the source are estimated using MIST stellar evolution models. This parametrization imposes astrophysical constraints that help identify the physically most probable solutions. We test our method on the complex microlensing event OGLE-2017-BLG-0114, 
which exhibits signatures that can be traced to the complexity of the source system. We successfully constrained the microlensing models, achieving improvements in  the Einstein ring radius estimates by up to an order of magnitude in the case of binary source models.
}

\maketitle
\section{Introduction}

Gravitational microlensing has emerged as a powerful technique for detecting and characterizing exoplanets. In the three decades since it was proposed as an exoplanet detection method by \cite{1991ApJ...374L..37M}, microlensing surveys have successfully discovered more than 270 planetary systems \footnote{\url{https://exoplanetarchive.ipac.caltech.edu}}. The method relies on the gravitational field of a foreground lens system magnifying the light of a background source star when the two become aligned along the line of sight. When the lens system contains a planetary companion, the planet can produce characteristic short-duration anomalies in the light curve, enabling the detection of planets with masses down to Earth-mass and below. Since the method does not rely on the light from the planet hosting star, it is particularly sensitive to planets at and beyond the snow line of their host stars, as well as to those located at distances from the Galactic center that are currently inaccessible by any other method \citep{2012ARA&A..50..411G, 2018Geosc...8..365T, Mroz2020}.

However, as gravitational microlensing is a highly nonlinear phenomenon, interpreting microlensing light curves is often complicated by degeneracies and additional effects that can mimic or obscure planetary signals.  One such effect is the xallarap effect which arises from the orbital motion of the source star in a binary system. This effect can introduce asymmetries in the light curve that must be carefully disentangled from genuine lens-system features. Stellar binary systems are common in solar neighborhood \citep{1969JRASC..63..275H, 2010ApJS..190....1R, 2024ApJS..271...55K} and we may expect that if a significant fraction of microlensing source stars are in binary systems, this effect should be present in many microlensing events. Indeed, \cite{2005ApJ...633..914P} found that out of 22 analyzed asymmetric microlensing events, approximately $23\%$ suffer from xallarap effect.

While the xallarap effect has been invoked to explain light curves in several microlensing events, its presence significantly increases the complexity of modeling and can lead to multiple degenerate solutions. The first detailed analyses of a planetary microlensing event suspected of exhibiting xallarap was OGLE-2007-BLG-368  \citep{2010ApJ...710.1641S}. In that case, the authors argued that the features of the light curve could only be explained by a specific binary lens–source configuration. However, since this model produced asymmetric residuals, they considered models including either parallax or xallarap effects, with the latter providing a better fit to the data. However, the authors noted that the consideration of this secondary effect did not affect the estimation of the microlensing parameters of the lens. In the case of MOA-2006-BLG-074 \citep{2021AJ....162...59R}, an anomaly in the light curve lacking clear caustic-crossing features was initially suspected to be caused by an additional mass in the lens system. However, this interpretation provided good fit to the data only after accounting for the orbital motion of the lens system. An alternative scenario considered was that the observed asymmetry was caused by a binary-source system undergoing orbital motion. Among these three-body models, the binary-source interpretation was favored by the authors, as it yielded a lower $\chi^2$ value and resulted in a more physically plausible interpretation. For another event without clear caustic crossings, KMT-2019-BLG-0414 \citep{2022A&A...666A.132H}, a binary-lens solution reproduced the data slightly better than the binary-source model with xallarap effect. However, since the difference in $\chi^2$ was minimal, both interpretations were considered viable by the authors. These two cases illustrate that for events without caustic crossings, both binary-lens and binary-source solutions warrant careful inspection.
The analysis of the planetary event OGLE-2017-BLG-0448 \citep{2024AJ....167..162Z} demonstrated that all four additional effects, parallax, xallarap, an additional lens, and an additional source, can be intertwined in complex often unresolved degeneracies. This was demonstrated in the systematic KMTNet Planetary Anomaly Search, which included the OGLE-2017-BLG-1777 event \citep{2024AJ....167...88R}. The light curve of this event was best explained by a model incorporating the xallarap effect, which led to a change in the binary-lens mass ratio by an order of magnitude compared to the static-source model.

One of the most interesting cases is the planetary microlensing event MOA-2010-BLG-328, whose analysis based on photometric data was performed by \cite{2013ApJ...779...91F}. They considered models that include orbital motion in either the source or the lens system. The event was later re-analyzed using high-resolution follow-up constraints by \cite{2025AJ....170..310V}. The authors could not find solutions from the original work that would satisfy these constraints. Moreover, they argued for a model of the event that includes all the previously considered effects: parallax, orbital motion of the lens and source, extended by the presence of additional luminous objects in the system. Considering two effects arising from the binary source system, additional flux and orbital motion, was well justified, as high-resolution imaging showed that \cite{2013ApJ...779...91F} underestimated the flux from the source system. MOA-2010-BLG-328 proved that modeling of microlensing source systems can be especially complex, and most importantly, that astrophysical properties of the source system can influence the microlensing event in complex  ways, and therefore should be taken into account.

In this paper, we introduce a new modeling framework, which we refer to as physical source parametrization. In this approach we directly sample the physical parameters of the source stars during the light-curve fitting process and connects them self-consistently to observable quantities through stellar evolution models. This allows astrophysical constraints to be incorporated directly into the Markov Chain Monte Carlo (MCMC) sampling, rather than applied a posteriori. The key advantage of this approach is that it naturally restricts the parameter space to physically plausible regions and provides direct access to correlations between microlensing observables and stellar properties. In particular, it enables significantly improved constraints on the angular Einstein radius and helps to resolve degeneracies that are otherwise difficult to break using standard techniques.

We demonstrate the performance of this method using the planetary microlensing event OGLE-2017-BLG-0114 as a case study. This event exhibits a complex light curve with multiple viable interpretations, making it a particularly suitable testbed for physically constrained modeling. The event was first presented by \citet{2021AcA....71....1P}, but since their preliminary models favored a close-orbit planet and their study focused on wide-orbit planet demographics, a comprehensive analysis of this event was not carried out. Our modeling approach will be made available through the \texttt{MulensModel}\footnote{\url{https://github.com/rpoleski/MulensModel}}  Python package \citep{2019A&C....26...35P}.

The structure of this paper is as follows. In Section 2, we describe the microlensing models used in this work, including single- and binary-lens configurations, the parallax effect, and source system models incorporating the xallarap effect. Section 3 introduces our astrophysical source parametrization approach and the relations between the physical parameters. Section 4 describes the observational data for OGLE-2017-BLG-0114. In Section 5, we present the modeling of the event using the standard parametrization, including the derivation of physical parameters. Section 6 presents the modeling using the astrophysical source parametrization, covering the initial states, fitting procedure, and results for both the source and lens systems. We discuss the method in Section 7, and summarize our conclusions in Section 8.

\section{Microlensing models}
\subsection{1L1S}
The simplest microlensing event occurs when a single point-mass lens magnifies a single background source (the 1L1S model) moving with rectilinear relative motion. In this case, the magnification is described by the \cite{doi:10.1126/science.84.2188.506} formula:
\begin{equation}
    A(u)=\frac{u^2+2}{u\sqrt{u^2+4}},
    \label{equ:magnif}
\end{equation}
where $u(t)$ is the projected separation between the source and the lens at time $t$, expressed in units of the angular Einstein radius $\theta_{\mathrm{E}}$.
This separation can be written as \citep{1986ApJ...304....1P}:
\begin{equation}
u(t)= \sqrt{u_0^2 + \tau(t)^2}, \qquad \tau(t)=\frac{t-t_0}{t_{\mathrm{E}}},
\label{equ:u}
\end{equation}
where $u_0$ is the impact parameter, $t_0$ is the time of closest approach, and $t_{\mathrm{E}}$ is the Einstein timescale.

To account for blended light, we express the observed flux as:
\begin{equation}
    f(t)=f_{\mathrm{S}} A(t) + f_{\mathrm{B}},
\end{equation}
where $f_{\mathrm{S}}$ is the unmagnified source flux and $f_{\mathrm{B}}$ represents any additional blended flux.

For each set of microlensing parameters, we first compute the magnification $A(t)$ and then determine $f_{\mathrm{S}}$ and $f_{\mathrm{B}}$ via linear regression \citep{1999ApJ...522.1037R}.  The resulting symmetric light curve is commonly known as the Paczyński curve.

To account for finite-source effects, we include an additional parameter $\rho$, defined as the angular radius of the source star normalized to the Einstein ring radius, $\rho = \theta_* / \theta_{\mathrm{E}}$.
\subsection{2L1S}
When the observed light curve exhibits more complex features than the standard Paczyński curve, this is often an indication of an additional mass component in the lens system. The binary-lens model (2L1S) requires extending the \cite{1986ApJ...304....1P} parameters by three additional ones. Two of these parameters describe the location of the secondary lens relative to the primary lens and the source trajectory. These parameters are typically: $s$ -- the projected separation between the primary lens and the companion relative to $\theta_{\mathrm{E}}$ and $\alpha$ -- the angle between the source trajectory and the binary axis. The third parameter describes the secondary lens mass and typically it is $q$ -- the ratio of the secondary lens mass and the primary lens mass. 

\subsection{Parallax effect}
The asymmetry of the light curve in a microlensing event can be introduced by additional motion in the observer, the lens, or the source systems. Typically, the most prominent of these is the parallax effect, caused by the motion of the observer due to the Earth's orbital motion. This motion results in an apparent deviation from the straight-line trajectory of the source relative to the lens. To account for the parallax effect, we include two additional parameters: the north ($\pi_{\mathrm{E},N}$) and east ($\pi_{\mathrm{E},E}$) components of the microlensing parallax vector $\boldsymbol{\pi}_{\mathrm{E}}$, defined in equatorial coordinates.

\subsection{Source system models}
In addition to parallax and the complexity of the lens system, microlensing light curves can also be influenced by the characteristics of the source system.
\subsubsection{1L2S}
The first possibility is an additional luminous source (1L2S). In this case, the effective magnification is given by the normalized sum of the magnifications of the two sources:
\begin{equation}
A=\frac{A_1 f_{\mathrm{S1}} + A_2 f_{\mathrm{S2}}}{f_{\mathrm{S1}} + f_{\mathrm{S2}}},
\label{equ:2SA}
\end{equation}
where $A_i,~i=1,2,$ denotes the lensing magnification of each source star with flux $f_{{\rm S}i}$. The magnifications $A_i$ are calculated analogously to Eqs.~\ref{equ:magnif} and \ref{equ:u}. Assuming that the transverse velocity of both sources relative to the lens is the same, the inclusion of the second source introduces three additional microlensing parameters  ($t_{\mathrm{0,2}}$, $u_{\mathrm{0,2}}$, $\rho_{\mathrm{2}}$) and one flux parameter ($f_{\rm S2}$).

\subsubsection{Xallarap effect}

The inverse of the microlensing parallax effect is the orbital motion of the source in a binary system, known as the xallarap effect \citep{1992ApJ...397..362G}. We first consider the case of a single luminous source, assuming that the companion is either very faint, unmagnified, or unresolved (i.e., its projected orbit is much smaller than both $\theta_{\mathrm{E}}$ and $u_0$).

The xallarap motion can be described using Keplerian orbital elements \citep{2024AJ....167..162Z}. In our parametrization, we adopt:
$\xi_\mathrm{P}$ -- the orbital period of the xallarap,
$\xi_\mathrm{a}$ -- the semi-major axis of the xallarap orbit as a fraction of the Einstein ring radius,
$\xi_\mathrm{i}$ -- the inclination,
$\xi_\mathrm{\Omega}$ -- the longitude of the ascending node;
$\xi_\mathrm{u}$ -- the argument of latitude at the reference epoch ($t_{0,\xi}$),
$\xi_\mathrm{\omega}$ -- the argument of periapsis,
and $\xi_\mathrm{e}$ -- the eccentricity.
We set the reference direction for the xallarap orbital parameters to align with the direction of relative lens--source proper motion, and we chose the plane of the sky as the reference plane.
To model how the xallarap effect influences the relative position between the lens and the source, we calculate the orbital motion of the luminous source around the system's center of mass using standard orbital integration.
We determine the source's position vector projected onto the reference plane at each epoch $\mathbf{r}_1(t)$, and at the xallarap reference time $\mathbf{r}_1(t_{0,\xi})$.
By subtracting them ($\mathbf{r}_1(t) - \mathbf{r}_1(t_{0,\xi})$) we obtain the xallarap shift. By construction, the xallarap shift at~$t_{0,\xi}$ vanishes, and the position of the center of mass of the source system relative to the source component is equal to~$-\mathbf{r}_1(t_{0,\xi})$. The resulting shift is then added to the relative positions of the source and the lens.

\subsubsection{1L2S + \XP }
In a realistic binary source system, the orbital motion responsible for the xallarap effect is inherently linked to the presence of a secondary object. In most cases, this companion is expected to be luminous, regardless of whether it is directly detectable in the data. We therefore relax the assumption that the secondary source contributes negligible flux and consider a model with two luminous sources undergoing orbital motion (1L2S + \XP). In this framework, the total observed flux is the sum of the magnified fluxes from both source components, each following the coupled orbital motion around the system's center of mass.

If we denote the position of the primary source relative to the center of mass of the source system as $\mathbf{r}_1(t)$, then the position of the secondary source is given by $- \mathbf{r}_1(t) / q_{\mathrm{source}}$, where $q_{\mathrm{source}}$ is the mass ratio of the two components of the system. In this case, the xallarap shift of the secondary source becomes $- \mathbf{r}_1(t)/ q_{\mathrm{source}} - \mathbf{r}_1(t_{0,\xi})$. Thus, extending the xallarap model to include a second luminous source requires only one or two additional parameters: the source mass ratio $q_{\mathrm{source}}$ and, if finite-source effects are considered, the normalized angular radius of the secondary source, $\rho_2$.
Despite introducing only a small number of additional parameters, this model significantly increases the flexibility of the fit due to the additional degrees of freedom associated with the second source.
At the same time, such configurations are astrophysically well motivated, as the probability of a microlensing event involving two unrelated luminous sources is extremely low, making a bound binary source the most plausible scenario.

\section{Astrophysical source parametrization}
Modeling of complex microlensing events often leads to multiple degenerate solutions that provide equally good fits to the observational data. This is especially true for models that include the xallarap effect. Another issue with the conventional methodology, based on incrementally adding microlensing parameters to describe increasingly complex models, is that it may overlook, at various stages of the analysis, whether the emerging solutions remain astrophysically plausible. As a consequence, the procedure may favor solutions that fit the data better at an intermediate stage, but do not ultimately lead to the correct astrophysical interpretation once the full model is considered.

\subsection{Concept}
To mitigate this issue, we decided to adopt an approach in which the physical parameters of the source stars are used directly in the MCMC sampling, instead of simply fitting the $f_{\mathrm{S}i}$ parameters via linear regression for each MCMC link and using the resulting posterior in post-mortem analysis to derive physical parameters. In this ``half-physical'' parametrization we introduced five additional fitting parameters. This approach allowed us to impose strong astrophysical constraints on quantities that are otherwise independent, such as $f_{\mathrm{S}i}$, $\rho_{i}$, $q_{\mathrm{source}}$, and the orbital parameters of the source system. These five additional fitting parameters are: the initial mass of the primary source ($M_{\mathrm{init,S}1}$), its equivalent evolutionary point \citep[$\mathrm{EEP}_{\mathrm{S}1}$ -- a monotonic index dividing physically significant stages of stellar evolution into equally spaced steps;][]{2016ApJS..222....8D}, the metallicity $\mathrm{[Fe/H]}_{\mathrm{S}1}$, the $V$-band extinction $A_{\mathrm{V}}$, and the source distance $D_{\mathrm{S}}$.

At each MCMC step, these parameters were used to compute the expected brightness of the source stars in all modeled passbands. To do so, we employed the MESA Isochrones and Stellar Tracks (MIST) v1.2 evolutionary tracks, hereafter referred to by the abbreviation ``\ET'' \citep{2016ApJ...823..102C, 2016ApJS..222....8D}, together with their interpolations implemented in the \texttt{brutus} algorithm \citep{2025arXiv250302227S}.

For single-source models, we used \texttt{brutus} to determine the stellar parameters based on the sampled values of $M_{\mathrm{init, S1}}$, $\mathrm{EEP}_{\mathrm{S}1}$, and $\mathrm{[Fe/H]_{\mathrm{S}1}}$. The algorithm then uses $D_\mathrm{S}$, $A_V$, and bolometric corrections  to obtain synthetic magnitudes for the source star in each observed passband. These bolometric corrections are derived from atmospheric models, with the computational cost reduced through neural-network–based grid mapping of the model grids. Synthetic magnitudes are then transformed to flux space and used as the values of $f_{\mathrm{S}i}$ in the microlensing model. The blending flux ($f_{\mathrm{B}}$) is the only remaining flux parameter, which is calculated.

In the case of binary-source models, we explicitly sampled the physical parameters only for the primary component of the system. The properties of the secondary source were then inferred from the sampled mass ratio $q_{\mathrm{source}}$, assuming that both components belong to the same binary system and therefore share identical age, metallicity, extinction, and distance. This strategy ensures full internal consistency of the binary’s physical parameters while reducing the dimensionality of the parameter space compared to modeling the two sources independently. 
If the models for the secondary source could not be obtained because its parameters fall outside the ranges covered by the MIST models, we fixed its flux to zero.
  
We note that the parameters of the sources used for generating synthetic magnitudes are theoretical and inferred using several simplifying assumptions, such as non-rotating stars, solar-scaled abundance patterns, and the total-to-selective ratio $R_{\mathrm{V}} = 3.1$, among others.

\subsection{Relations between parameters}
When we considered the binary-source model with this parametrization, we were able to obtain information on $\theta_{\mathrm{E}}$ from two independent lines of evidence. 
Firstly, since $D_{\mathrm{S}}$ is one of the sampled parameters, we can calculate $\theta_{\mathrm{E}}$ using the value of the relative semi-major axis $\xi_{a}$ and the physical semi-major axis $a$ of the source system, derived from Kepler’s third law:
\begin{equation}
\theta_{\mathrm{E, orbit}} = \frac{a} {D_{\mathrm{S}} \xi_{a}},
\end{equation}
with
\begin{equation}
a = (1+ q_{\mathrm{source}})^{1/3} \left(\frac{M_{\mathrm{S}1}}{M_\odot}\right)^{1/3}\left(\frac{\xi_{P}}{\mathrm{{yr}}}\right)^{-2/3}\mathrm{AU},
\end{equation}
where $M_{\mathrm{S1}}$ is the current mass of the primary source.

Secondly, we can use the information on the physical radius of the source, $R_{*,\mathrm{S1}}$, from the MIST models, together with the modeled relative angular size of the source:
\begin{equation}
\theta_{*,\mathrm{S1}}= \left(\frac{R_{*,\mathrm{S1}}}{\mathrm{R_\odot}}\right) \left(\frac{R_{\odot}}{\mathrm{AU}}\right)  \left(\frac{D_{\mathrm{S}}}{\mathrm{kpc}}  \right)^{-1} \mathrm{mas}
\end{equation}
\begin{equation}
\theta_{\mathrm{E, size}} = \theta_{*,\mathrm{S1}} / \rho_{1}
\label{equ:theta_size}
\end{equation}
The relative angular size of the secondary source $\rho_2$ is not a free parameter of the sampling; instead, it is fixed to the value calculated based on $R_{*,\mathrm{S2}}$ from the MIST models and the obtained $\theta_{\mathrm{E, size}}$, ensuring self-consistency.
\begin{equation}
\rho_{2}=\theta_{*,\mathrm{S2}} / \theta_{\mathrm{E, size}} = \left(\frac{R_{*,\mathrm{S2}}}{\mathrm{R_\odot}}\right) \left(\frac{R_{\odot}}{\mathrm{AU}}\right)  \left(\frac{D_{\mathrm{S}}}{\mathrm{kpc}}  \right)^{-1}  \left(\frac{\theta_{\mathrm{E, size}}}{\mathrm{mas}}  \right)^{-1}  
\end{equation}
In the sampling, we imposed a likelihood penalty for inconsistent values of $\theta_{\mathrm{E, orbit}}$ and $\theta_{\mathrm{E, size}}$ in a form:
\begin{equation}
\mathcal{L}(\theta_{\mathrm{E, size}}) = \mathcal{N}(\mu=\theta_{\mathrm{E, orbit}}, \sigma= 0.001) \bigl(\theta_{\mathrm{E, size}}\bigr),
\label{equ:prior_theta}
\end{equation}

\section{OGLE-2017-BLG-0114 observational data}

The microlensing event OGLE-2017-BLG-0114 was alerted by the Optical Gravitational Lensing Experiment (OGLE) on 2017 February 16, Heliocentric Julian Date (HJD) $\sim2457801$. The detection was made by the Early Warning System (EWS\footnote{\url{https://ogle.astrouw.edu.pl/ogle4/ews/ews.html}}; \citealt{2003AcA....53..291U}) in the data from the 1.3-m Warsaw Telescope. The telescope is located at the Las Campanas Observatory in Chile and is equipped with the $1.4~\mathrm{deg}^2$ field of view mosaic CCD camera \citep{2015AcA....65....1U}. The  event was located at the equatorial coordinates $(\mathrm{RA},\mathrm{Dec})_{J2000}$ = (17:21:57.74,~-29:37:25.3) or Galactic coordinates $(l,b) = (356.^\circ62,~3.^\circ96)$, placing it within the BLG615 field of the OGLE-IV survey (the fourth phase of OGLE). This field was typically observed once every two nights. OGLE images were mostly taken in the $I$-band, with occasional observations in the $V$-band every few days.
The OGLE data were reduced using a variant of difference image analysis (DIA; \citealt{1996ApJ...473L..87C, 1998ApJ...503..325A}) optimized by \cite{2000AcA....50..421W} and \cite{2008AcA....58...69U}. Photometric extraction was done using reference images taken one year after the event, in 2018.

Near the peak, the light curve of the event visibly deviated from the standard symmetric model by \cite{1986ApJ...304....1P}. 

One year after the main magnification, the OGLE survey observed an additional brightening of the event. This secondary peak started on April 8, 2018 (HJD $\sim 2458217$), lasted over the course of 21 nights, and showed an amplitude of $\sim0.1$~mag. To this day, no further brightening of the target has been observed by the OGLE survey.

\begin{figure}
    \centering
    \includegraphics[width=1\linewidth]{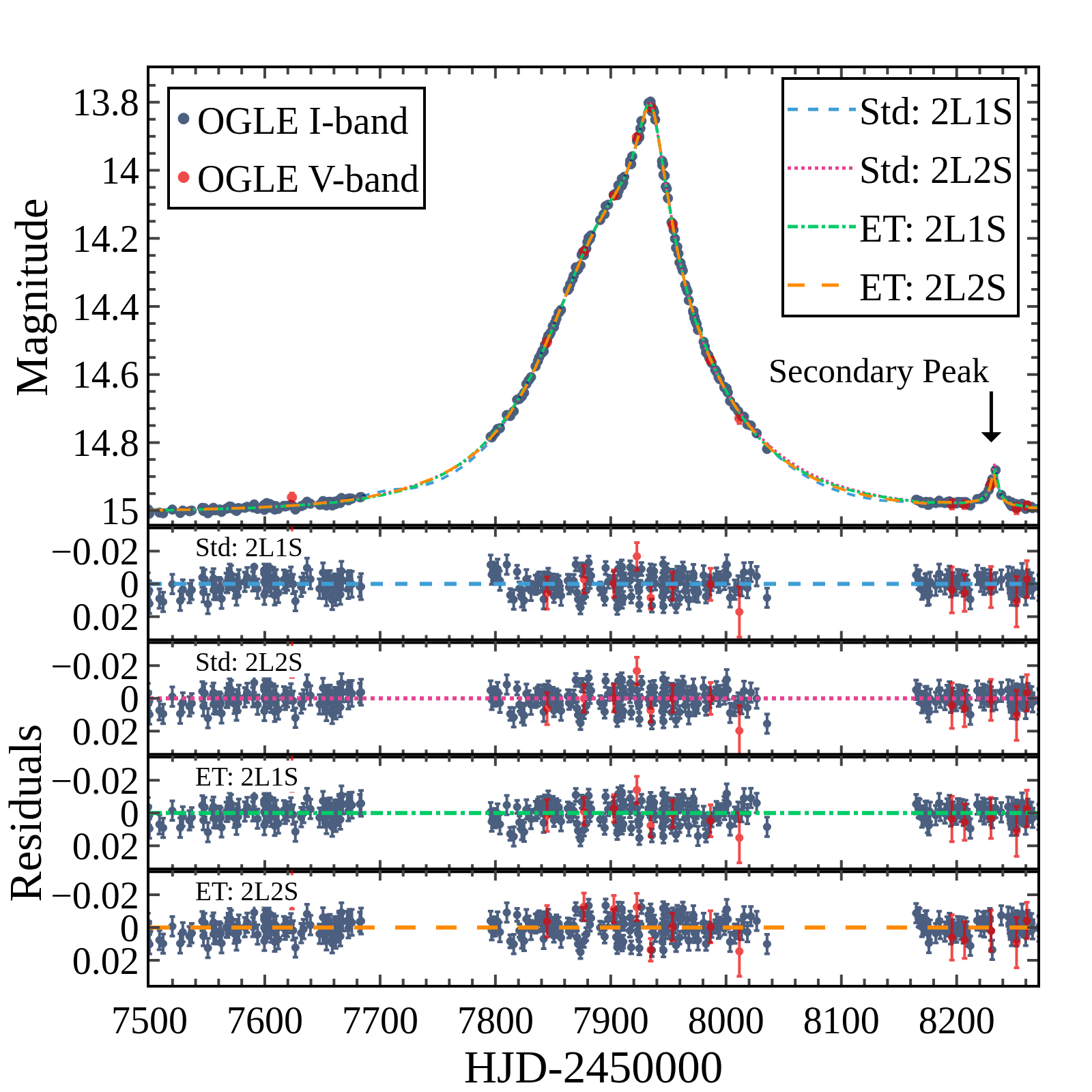}
    \caption{Light curves of the microlensing event OGLE-2017-BLG-0114. All solutions include the xallarap effect.}
    \label{fig:model_ET}
\end{figure}

\begin{figure}
    \centering
    \includegraphics[width=0.85\linewidth]{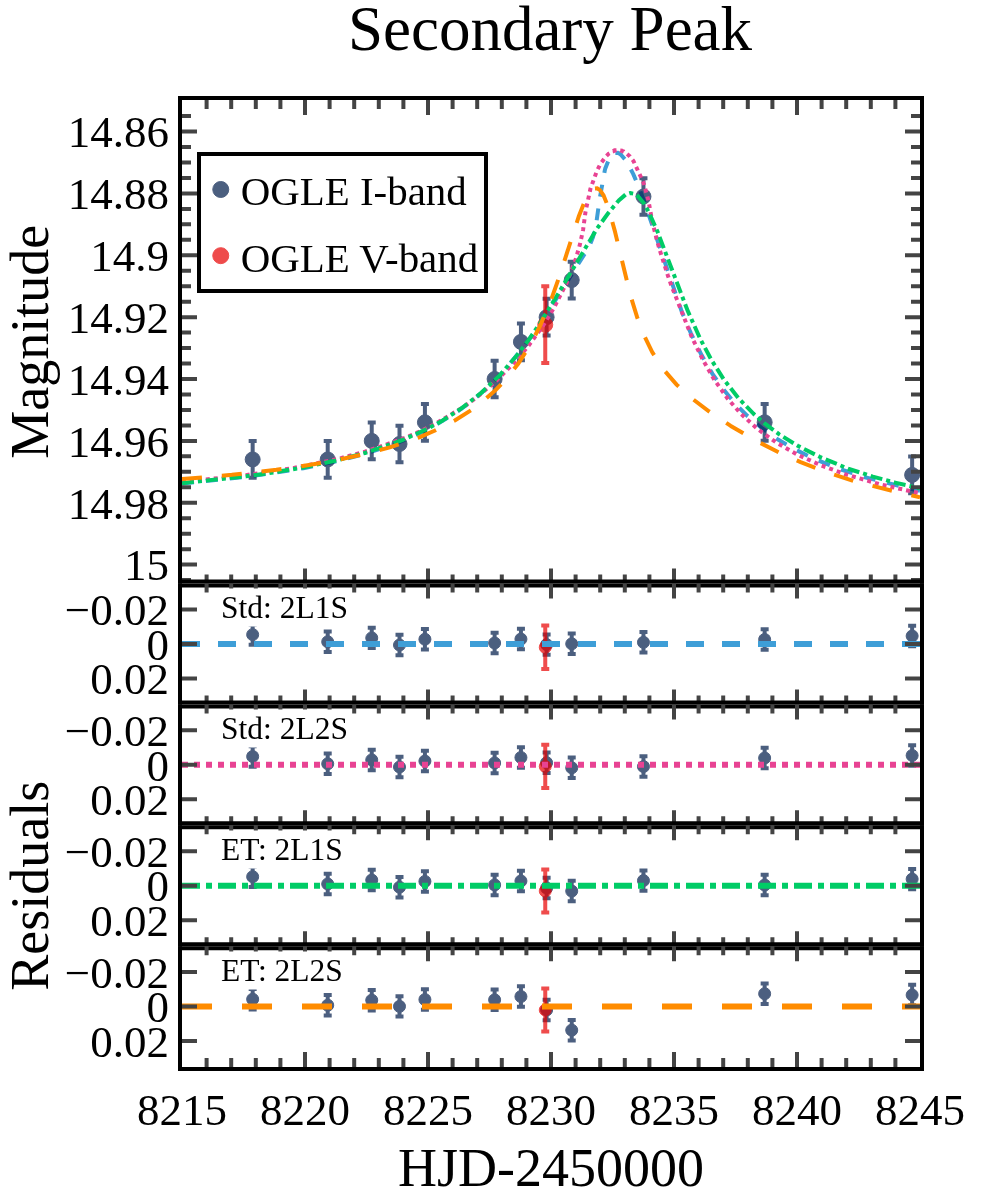}
    \caption{Light curves of the planetary anomaly in OGLE-2017-BLG-0114. All solutions correspond to those presented in Figure~\ref{fig:model_ET}.}
    \label{fig:model_zoom_ET}
\end{figure}

The event’s light curve is presented in Figure~\ref{fig:model_ET}, with a detailed view of the secondary peak shown in Figure~\ref{fig:model_zoom_ET}. 

Although additional data for this event were obtained by other facilities (Spitzer, the DECam Plane Survey, Gaia, and ATLAS), we do not include them in the present analysis, as our goal is to illustrate the methodological framework rather than to provide a comprehensive modeling of the event.  These data sets either provide incomplete coverage of the event or lack sufficient photometric precision.

We calibrated the OGLE $I$- and $V$-band measurements to ensure that the reported magnitudes are in the standard $I$ (Cousins) and $V$ (Johnson) pass bands. The calibration was performed using offsets derived from observations of standard stars obtained by the OGLE project, following the procedure described by \citet{2008AcA....58...69U,2011AcA....61...83S,2015AcA....65....1U}. 

Photometric reduction pipelines can have difficulty estimating the impact of systematics in the data; as a result, uncertainties in the raw measurements are often underestimated and display broader distributions than expected for a Gaussian scatter. When fitting models to such data sets, underestimated error bars can lead to underestimated uncertainties in the resulting parameters. To renormalize photometric error bars, we scaled them using the formula for the $i$-th data point \citep{2012ApJ...755..102Y}: 
\begin{equation}
\sigma_{\mathrm{new}, i}= k_j \sigma_i
\label{equ:err}
\end{equation}
where $\sigma_i$ is an initial error bar, $k_j$ is the renormalization coefficient of the $j$-th data set. 

Instead of fitting the renormalization coefficients, we could use the coefficients reported by \citet{2016AcA....66....1S}. We decided not to do that as those were based on reductions performed with earlier OGLE-IV reference images and could be inaccurate. 
Instead, we determined the coefficients directly from the event data. We fixed the microlensing parameters to the values obtained from the preliminary best fit and treated the scaling parameters $k_j$ for the $I$- and $V$-band observations as the only free parameters. These were sampled using the Markov Chain Monte Carlo (MCMC) method with the likelihood function:
\begin{equation}
\ln \mathcal{L} = -\frac{1}{2} \sum_{i=1}^{N} \ln \left( 2 \pi \sigma_{\mathrm{new}, i}^2 \right).
\end{equation}
The inferred renormalization coefficients, $k_{\mathrm{OGLE~I-band}}= 1.98$ and $k_{\mathrm{OGLE~V-band}}= 1.39$, were applied to rescale the OGLE photometric uncertainties prior to the substantial model fitting. All reported $\chi^2$ values are calculated using the rescaled uncertainties.

\section{OGLE-2017-BLG-0114 modeling using standard parametrization (Std)}

To provide a benchmark for our method, we first fitted 2L1S + \PX~ + \XP~ and 2L2S + \PX ~+ \XP~ models to the data using the standard microlensing parametrization. 

All microlensing modeling in this work was performed using the \texttt{MulensModel} Python package \citep{2019A&C....26...35P}, together with the affine-invariant MCMC ensemble sampler \citep{2010CAMCS...5...65G} implemented in the \texttt{emcee} code \citep{2013PASP..125..306F}.
For each MCMC modeling run, the number of chains and steps was adjusted to ensure convergence. All runs with fixed error-bar scaling parameters achieved mean acceptance fractions between 0.019 and 0.106, and the number of steps was at least 25 times greater than the mean autocorrelation time; no thinning was applied (for definitions, see \citealt{2013PASP..125..306F}).
The binary-lens magnification was calculated using the \texttt{VBBinaryLensing} software \citep{2010MNRAS.408.2188B} or its further optimized version, the \texttt{VBMicrolensing} software \citep{2025A&A...694A.219B}. These calculations require solving fifth-order complex polynomial efficiently, which was done using the \citet{2012arXiv1203.1034S} algorithm.

\subsection{Main peak analysis}
The main and secondary peaks are well separated, which allowed us to begin our analysis by fitting models to the data without the secondary peak.
We used five years of observations of the event, spanning 2015 to 2019, excluding 48 measurements over 67 nights taken during the secondary peak.

\subsubsection{1L1S + \PX~+ \XP}

Xallarap models tend to be affected by significant degeneracies. Since we initially excluded data points around the secondary peak and did not have a reliable static 1L1S model solution to start with, we adopted a step-by-step fitting approach. This allowed us to thoroughly explore the parameter space and avoid missing solutions that could have become preferred once the secondary peak data were included.

We initially explored the parameter space by running MCMC sampling with the walkers’ initial positions drawn from uniform distributions: between 0 and 360 degrees for the angles $\xi_{\mathrm{\Omega}}$ and $\xi_{\mathrm{u}}$, and between 0 and 180 degrees for $\xi_{\mathrm{\omega}}$ and $\xi_{\mathrm{i}}$. The initial values of $\xi_{\mathrm{a}}$ were drawn from a log-uniform distribution in the range (0.001, 0.1). For the period of the xallarap orbit, $\xi_{\mathrm{P}}$, we performed a grid search by running 20 independent MCMC chains, each exploring a different period range: $(\frac{3}{4} \xi_{\mathrm{P},i}, \frac{4}{3} \xi_{\mathrm{P},i})$, where $\xi_{\mathrm{P},i}$ represents elements of a geometric series of ten values spanning from 20 to 1000 days \citep{2024AJ....167..162Z}.

To efficiently and reliably handle the multiple solution modes found in the MCMC posterior, we decided to employ a clustering algorithm. After testing the performance of several available methods, we selected the Skinny-Dip algorithm \citep{10.1145/2939672.2939740}, which performs a parameter-free iterative test of unimodality. Compared to other clustering algorithms, it is highly noise-robust and identifies only separated groups of solutions. We performed a cut on the posterior by selecting models with $\chi^2$ statistic lower than the lowest $\chi^2$ value obtained from the model increased by half the number of degrees of freedom ($min(\chi^2)  + d.o.f. /2  = 667+ 611 /2 = 972.5$). We then applied the Skinny-Dip algorithm to the remaining posteriors from all twenty different orbital period ranges. As a result, we identified 125 clusters: 66 corresponding to positive $u_0$ values and 59 corresponding to negative ones. From each cluster, we selected the model with the lowest $\chi^2$.  We then ran 125 new samplings with initial walker positions drawn from narrow Gaussian distributions centered on these best-fit models. We then compared the results by checking whether the parameters of the lowest-$\chi^2$ model from one sampling fell within the $\pm1.5\sigma$ range of the posteriors from the other sampling. As a result, we retained ten unique models with positive $u_0$ and eight with negative $u_0$. We further excluded models with $\chi^2 > \min(\chi^2) + 15$ for each of $u_0$ ranges, $\xi_{\mathrm{e}} > 0.95$, and with orbital periods $350 < \xi_{\mathrm{P}} < 380$ days, as these solutions are highly degenerate with parallax. The remaining one model with positive $u_0$ and three models with negative $u_0$ were subsequently used as starting points for the analysis including both peaks.

\subsubsection{1L2S + \PX~+ \XP}
To find the correct solutions for the xallarap orbit, we repeated the steps previously applied to the single-source model. First, we considered models without the finite-source effect. We performed a period grid search, as described above, but this time with initial values of $q_{\mathrm{source}}$ drawn from a normal distribution with $\mu = 1$ and $\sigma = 0.5$.
In the resulting posteriors, using the clustering algorithm, we identified 754 groups of model solutions for positive $u_{0,1}$ and 617 for negative $u_{0,1}$. Since the trajectory of the secondary source is fully determined by its orbital motion, this effectively breaks the degeneracy between positive and negative values of the secondary impact parameter $u_{0,2}$.
In the second step, we ran the MCMC sampling starting from each of the 1371 models. In the results, after repeating the same cuts as for 1L1S + \PX~+ \XP~ models, we identified 23 unique, well-converged solutions: 14 with positive $u_{0,1}$ and nine with negative $u_{0,1}$. The periods of xallarap orbits in these models span from approximately  210 to 714 days, and the resulting $\chi^2$ values range from 661.4 to 681.0.

\subsection{Both peaks analysis}
The secondary peak visible in the light curve around HJD~$\sim 2458320$ can be explained either by an additional passing of one of the sources by the primary lens or by the presence of an additional planetary-mass object in the lens system.

In the first scenario, if the asymmetry of the main peak is caused by the xallarap motion of the source, the secondary peak could be due to either one of the sources involved in the xallarap motion or due to a third, additional source. In the first case, the source's orbit would have to be very precisely aligned to counteract its proper motion and approach the lens again, but none of the xallarap solutions we tested supported such a configuration. The alternative explanation involving a triple-source system is also unlikely, as a compact (in projection) triple system is improbable, and the presence of an unrelated third source is even less plausible. Another argument against the scenario in which both peaks are caused by the same lens is the drastically different duration of the two peaks. The duration of the secondary peak is roughly 100 times shorter than that of the main peak, whereas we would expect them to be comparable. Nevertheless, we attempted to fit a model based on this scenario, but it failed to reproduce the secondary peak. For these reasons, we excluded this scenario from further analysis.
\subsubsection{Lens companion}
We ultimately considered the second scenario, in which the secondary peak is caused by an additional object in the lens system. The ratio of durations of both peaks suggests the mass-ratio in the planet-to-star range. We started by examining all possible degenerate solutions. The first degeneracy we considered is the mentioned binary degeneracy, involving positive and negative values of $u_0$ ($u_0{+}$/$u_0{-}$). Next is the well-known close/wide degeneracy \citep{1999A&A...349..108D, 2005MNRAS.356.1409A}, where configurations with the planet located inside the Einstein ring ($s < 1$, close) and outside the Einstein ring ($s > 1$, wide) can produce similar light curves, especially when the source trajectory passes close but does not cross the planetary caustic. Another pair of solutions arises from the inner/outer degeneracy, which is related to the fact that the source trajectory can cross the binary axis either between the planetary and central caustic (inner) or beyond the planetary caustic (outer). This results in the source passing either close to the upper or lower cusp of the planetary caustic relative to the binary axis. Lastly, in the case where two sources are present in the system, either one of them can pass near the planetary caustic, leading to an additional pair of degenerate solutions (indicated as {\sP} and \sS, respectively). The list of all degeneracies is presented in Table~\ref{tab:deg}.

\begin{table}
\caption{Considered degeneracies $^a$.}
\label{tab:deg}
\centering
\begin{tabular}{c}
\hline\hline
$u_0{+}$ / $u_0{-}$ \\
\XP~orbit (denoted by numbers)\\
close / wide \\
inner / outer \\
\hline
For 2S models only:  \\
 \sP~/ \sS \\
\hline
\end{tabular}
\tablefoot{
\tablefoottext{a}{For an explanation, see Section~5.2.1}
}
\end{table}

To better control and sample the degenerate solutions listed above using MCMC, we initially reparametrized the traditional binary-lens parameters $s$ and $\alpha$. We combined known relations between the location of the planetary caustics and the parameters $s$ and $q$ \citep{2006ApJ...638.1080H}, along with information about the position of the source at a specific time near the secondary peak,  $t_{0, \mathrm{pl}} = 2458233$ HJD, to sample $s$ and $\alpha$ using two polar parameters ($\delta r$, $\delta \phi$). These  parameters describe the relative position of the source and the approximated position of the planetary cusp at $t_{0, \mathrm{pl}}$. Details of this reparametrization are provided in Appendix B.
In the case of the 2L1S + \PX~+ \XP~models, this approach resulted in four solutions for each xallarap orbit, yielding a total of 16 samplings. For the 2L2S + \PX~+ \XP~models, there were eight solutions for each xallarap orbit, resulting in a total of 184 samplings. These samplings were used to establish the limb-darkening coefficients following the procedure described later in Section 5.2.2.  Models with the lowest $\chi^{2}$ from each sampling were then adopted as starting points for the final microlensing fits, which used the standard parameters $s$ and $\alpha$ as free parameters and included the limb-darkening effect. The resulting models with the lowest $\chi^{2}$ values for the 2L1S~+~\PX~+~\XP~ and 2L2S~+~\PX~+~\XP~ configurations are presented in Table~\ref{tab:mulens_ET} (Std: 2L1S and Std: 2L2S, respectively).
\subsubsection{Physical parameters}
Based on the results of the microlensing modeling, both the source star and the blend are most likely typical Galactic bulge red giants. The physical parameters of the lens system were derived following the method of \citet{2004ApJ...603..139Y}, implemented as described in detail in \citet{2025A&A...698A.126M}.   All values reported in this section correspond to the best-fitting 2L1S + \PX~+ \XP~ model fitted using standard parametrization (Std: 2L1S). The results for this and the other models presented in Table~\ref{tab:mulens_ET} are summarized in Table~\ref{tab:parms_lens_ET}.  First, we estimated extinction and reddening values ($A_{\mathrm{I}}$, $A_{\mathrm{V}}$, $E(V-I)$) using the OGLE-IV photometric maps and methodology from \cite{2013ApJ...769...88N}. 
We selected red clump giant stars within $3'$ of the microlensing event position, excluding those in a region of visibly higher extinction, and adopted their mean $I$-band magnitude and $(V-I)$ color as reference values.
We found 
\begin{align}
A_{\mathrm{I}} &=1.837\pm0.038~\mathrm{mag},\\ A_{\mathrm{V}} &= 3.39\pm0.10~\mathrm{mag}, ~\mathrm{and}\\E(V-I) &=1.55\pm0.09~\mathrm{mag}.
\end{align}
We combined these estimates with the posterior distribution of the source flux from the microlensing modeling to derive the dereddened brightness and color of the source star. For solution~H, we found:
\begin{equation}
    I_{\mathrm{0}}=13.66\pm0.20~\mathrm{mag} \quad\mathrm{and}\quad (V-I)_{\mathrm{0}}=1.20\pm0.09~\mathrm{mag}.
    \label{equ:intr}
\end{equation}
To estimate the source angular size more precisely, we converted its dereddened color $(V-I)_0$ to $(V-K)_0$ using the color–color relations from \cite{1988PASP..100.1134B}. The angular radius, corrected for limb darkening, was then calculated following \cite{2018MNRAS.473.3608A}:
\begin{equation}
\begin{aligned}
    \log \theta_{*,\rm{LD}} = (0.562\pm0.009) + (0.051\pm0.003)~(V - K)_0 - \\ 0.2~K_0 -\log{2}.
    \label{equ:theta}
\end{aligned}
\end{equation}
\begin{equation}
\theta_{*,\rm LD} = 9.4\pm1.0  ~\mu\mathrm{as}.
\end{equation}
We assumed that, in this relation as well as in the following ones (Equations~\ref{equ:teff} and~\ref{equ:logg}), the posterior distribution of a given quantity is approximated as a normal distribution.
To determine the stellar parameters and corresponding limb-darkening coefficients, we first estimated the effective temperature using the empirical color relations for cool giants from \cite{2000AJ....119.1448H}:
\begin{equation}
\begin{aligned}
T_{\mathrm{eff}} = (8556.22 \pm 204.27) - (5235.57 \pm 352.83) ~(V-I)_0 + \\\quad (1471.09\pm 148.20)~(V-I)_0^2 =4406^{+530}_{-550}~ \rm K.
\end{aligned}
\label{equ:teff}
\end{equation}
The surface gravity was estimated using the relation for G–K giants and subgiants from \cite{1994AstL...20..755B}, assuming solar metallicity:
\begin{equation} 
\begin{aligned}
    \log{g} &= (8.0\pm0.04)~T_{\rm{eff}} + (0.31\pm0.04)~ M_{\rm{0,V}} +\\&(0.27\pm0.11)~\rm[Fe/H]-(27.15\pm2.1) = 2.3\pm2.1,
\end{aligned}
\label{equ:logg}
\end{equation}
with absolute brightness $M_{0,V} =0.2\pm0.2$ mag. The obtained limb-darkening coefficients are the following:
\begin{equation}
u_V = 0.80 \quad u_I = 0.57   \quad u_L=  0.24.
\end{equation}
We derived the angular Einstein radius from the scaled source radius obtained through finite-source effect modeling and the estimated angular source radius:
\begin{equation}
    \theta_{\rm{E}}= \frac{ \theta_{*,\rm{LD}}}{\rho}=0.86^{+1.56}_{-0.49} ~\mathrm{mas}.
\end{equation}
For the 2S models, the reported value is the average of the two Einstein radii, calculated separately for each source: $\theta_{\rm{E}} = \frac{1}{2}\left(\frac{ \theta_{*,\rm{LD},1}}{\rho_1} + \frac{ \theta_{*,\rm{LD},2}}{\rho_2} \right)$.
Combining this measurement with constraints on the microlensing parallax allowed us to determine the masses of the host and its companion \citep{2000ApJ...542..785G}:
\begin{equation}
M_{\rm{h}}=\frac{\theta_{\rm{E}}}{\kappa\pi_{\mathrm{E}}}=0.50^{+0.90}_{-0.28}~  \mathrm{M_{\odot}} \quad\mathrm{and}\quad  M_{\rm{c}}=qM_{\rm{h}} =11.5^{+21.7}_{-6.1} ~\mathrm{M_{\rm{J}}},
\end{equation}
where $\kappa \equiv 4 G/(c^{2} \mathrm{au}) \simeq 8.144~\rm{mas~M}_\odot^{-1}$.
The source--lens relative parallax is given by
\begin{equation}
 \pi_{\mathrm{rel}}=\pi_{\rm{E}}  \theta_{\rm{E}}= 0.18^{+0.33}_{-0.10}~\mathrm{mas},
\end{equation}
and the relative proper motion by:
\begin{equation}
    \mu_{\rm{rel}}= \frac{ \theta_{\rm{E}}}{ t_{\rm{E}}}= 1.8^{+3.3}_{-1.0}~\mathrm{mas~yr^{-1}}.
\end{equation}
Assuming a typical Galactic bulge distance for the source star of $D_{\rm S} = 8.54$~kpc \citep{2012ApJ...750..169P}, we determined the lens distance as
 \begin{equation}
    D_{\rm{L}}=\frac{\mathrm{au}}{ \pi_{\rm{rel}}+\pi_{\rm{S}}}=3.3\pm1.8 ~\mathrm{kpc},
\end{equation}
where $\pi_{\rm{S}}=\mathrm{au}/D_{\rm{S}}$. The projected semi-major axis of the secondary lens is then:
\begin{equation}
a_{\mathrm{L},\perp} = \theta_{\mathrm{E}}sD_{\rm L} =0.97\pm0.32~ \mathrm{au}.
\end{equation}
Under the same assumption for the source location, we can estimate the radial-velocity semi-amplitude of the source star implied by the xallarap orbit as
\begin{equation} 
    K_{\mathrm{RV,S}} = \frac{2 \pi  a_{\rm S} \sin\xi_{\mathrm{i}} }{\xi_{\mathrm{P}} \sqrt{1-{\xi_{\mathrm{e}}^2}}} = 44^{+96}_{-25}~ \mathrm{km/s},
\end{equation}
where the orbital semi-major axis in physical units is equal to:
\begin{equation}
a_{\rm S} = \theta_{\mathrm{E}}\xi_{\mathrm{a}}D_{\rm S} =1.46^{+2.85}_{-0.81} ~ \mathrm{au}
\end{equation}

\section{OGLE-2017-BLG-0114 modeling using astrophysical source parametrization (ET)}

\subsection{Initial states}
As the starting point for the sampling, we used all model solutions with \PX, \XP~ single-lens, and either one or two sources obtained from the standard parametrization analysis of the main peak. Based on the position of the source on the CMD (Fig.~\ref{fig:CMD_ET}), we assumed that at least one star in the system is a red giant. We then generated initial sets of physical parameters drawn from the distributions summarized in Table~\ref{tab:init}. If a set of drawn parameters could not produce valid stellar properties from the MIST models, we made an attempt with an inverted mass ratio, which means the assumption that at least the secondary source is a red giant. 
\begin{table*}[h!]
\tiny
\centering
\caption{Distributions of the sampled physical parameters.}
\label{tab:init}
\begin{tabular}{c|c|c}
\hline\hline         
\text{Parameter} & \text{Distribution} & \text{Bounds} \\
\hline        
$M_{\rm init, S1}$ & uniform & $0.08\,M_\odot$  to $100\,M_\odot$ \\
$[\mathrm{Fe/H}]_{\mathrm{S}}$& truncated normal & $\mu = 0.1$, $\sigma = 0.5$, from $-4.0$ to $0.5$ \\
$A_V$ & truncated normal & $\mu = 3.38^{\mathrm{a}}$, $\sigma = 0.1$, from $0.0$ to $5.0$ \\
$D_{\rm S}$ & Galactic model by \cite{2021ApJ...917...78K} & see Table \ref{tab:gal_prior} \\
$\text{EEP}_{\mathrm{S}1}$ & truncated normal & $\mu = 650^b$, $\sigma = 10$, from 10 to 808 \\
\hline  
For binary-source model & & \\
\hline
$q_{\mathrm{source}}$ & truncated normal & $\mu$ -- taken from analysis in standard parametrization, $\sigma = 0.001$, from $0.0$ to $100$ \\
\hline 
\end{tabular}
\tablefoot{
\tablefoottext{a}{From the analysis of red clump stars within 3' around the source; for description of this approach, see \cite{2025A&A...698A.126M}.}
\tablefoottext{b}{Value for red clump star.}
}
\end{table*}
\begingroup
\renewcommand{\arraystretch}{2.5}
\begin{table}
\caption{Settings of the Galactic model \citep{2021ApJ...917...78K}  used in simulating microlensing events for $D_\mathrm{S}$ prior.} 
\label{tab:gal_prior}
\centering 
\begin{tabular}{c|l}
\hline\hline Setting & Explanation\\
\hline 
\makecell{$v_{\oplus,\mathrm{N}} =-2.74~\mathrm{km~s^{-1}} $\\ $v_{\oplus,\mathrm{E}} =  28.11 ~\mathrm{km~s^{-1}}$ }&
\makecell[l]{Earth velocity at\\ $\mathrm{BJD_{TT}}=2457933.5$ \\projected on  the  plane of \\sky towards event's \\coordinates.}\\
$35~\mathrm{d} < t_{\mathrm{E}} < 200 ~\mathrm{d} $ &\makecell[l]{The Einstein crossing time.}\\
$15.0~\mathrm{mag} < I_{\mathrm{S}} <  17.5~\mathrm{mag} $   &  \makecell[l]{Source OGLE $I$-band \\brightness. \\See Section 5.2.2. }\\
$2.0~\mathrm{mag}<(V-I)_{\mathrm{S}}<3.1~\mathrm{mag}$ &\makecell[l]{ $(I-V)$ color index\\ of the source. }\\
$A_{\mathrm{I,RC}} = 1.83~\mathrm{mag} $&\makecell[l]{ Mean red clump extinction \\in the target field$^a$.}\\
$E(V-I)_{\mathrm{RC}} = 1.55 ~\mathrm{mag}$  &\makecell[l]{ Mean red clump reddening\\ in the target field$^a$.}\\
\hline 
\end{tabular}
\tablefoot{
\tablefoottext{a}{From the analysis of red clump stars within 3' around the source.}
}
\end{table}
\endgroup
If, after 1000 attempts per MCMC walker, we were still unable to generate the required number of valid states, we considered the tested value of $q_{\rm S}$ invalid and discarded the model.  
In this context, ``invalid'' means that the system would not correspond to either two red giants or a red giant paired with a main-sequence star. We assumed that systems consisting of a red giant with a stellar remnant ($q_{\rm S}$ resulting in too massive $M_{\mathrm{init,S}i}$) or a brown dwarf (too small $M_{\mathrm{init, S} i}$) are considered in single bright source models fitting, since the light from such objects would likely be below the detection limit.

All other microlensing model parameters were drawn from normal distributions with small variances, centered on the values identified in the standard analysis of the main brightening peak. Since we include finite-source effects at this stage, unlike in the standard analysis, we set the mean value of the primary source size parameter to $\rho_1 = 0.001$.
Out of the four considered 1L1S model solutions, we were able to generate physically valid initial states for all of them, whereas for the 1L2S models, only 13 out of 23 solutions yielded physically valid initial states.

\subsection{Fitting}
\subsubsection{Main peak analysis}
We began the model fitting with the physical parametrization of the source in a manner analogous to the standard analysis. First, we fitted the 1L1S~+~\PX~+~\XP~ and  1L2S~+~\PX~+~\XP~ models to the primary peak only, excluding all data from the secondary peak. 

In principle, any parameter listed in Table~\ref{tab:init} could be constrained by additional astrophysical priors during sampling. After testing several options, we found that applying an additional astrophysical prior only to the $D_{\mathrm{S}}$ parameter provides sufficient constraints. As for the initial state, we constructed this prior on $D_{\mathrm{S}}$ based on the Galactic model of \cite{2021ApJ...917...78K}, using the settings listed in Table~\ref{tab:gal_prior}.

For the 1S models, all four sampling schemes produced satisfactory results; however, the inferred values of $M_{\rm init, S1}$ were higher than expected for a typical Galactic bulge red clump star ($2.1 \le M_{\rm init, S1} \le 2.8~M_\odot$  for the models with the lowest $\chi^2$). In the case of the 2S models, the sampling resulted in more pronounced changes compared to the standard parametrization. Half of the samplings inferred an extremely distant main-sequence source system with very high reddening, in which the ratio of source fluxes to blending flux is $\sim 1/10$. As these solutions are physically implausible, we exclude them from further analysis. Sampling that led to the lowest $\chi^2$ statistic ($\chi^2=671.6$) yielded more physically justified sources' parameters, consistent with those expected for  far-side Galactic bulge red clump stars. The remaining eleven sampling runs yielded significantly higher $\chi^2$ values ($\chi^2 > 805.2$). We ultimately left ten 2S model solutions that were used in the analysis of the full event. We note that we proceed with all ten solutions only as an insurance measure when using a newly introduced  method. Otherwise, selecting only the model with the lowest $\chi^2$ would be justified and would substantially reduce the computational resources required compared to the standard analysis.

\subsubsection{Both peaks analysis}
In the next step, we included the observations of the secondary peak and performed the fitting in the same manner as in the standard analysis of binary-lens models starting from eleven solutions from main peak analysis. We began by reparameterizing $(s, \alpha) \longrightarrow (\delta r, \delta \phi)$ to efficiently sample all degenerate solutions. For each set of sampled solutions, we selected those with the lowest $\chi^2$. These were then used as starting points for fitting in the original $(s, \alpha)$ parametrization, including limb-darkening effects. The limb-darkening coefficients for each solution were assigned based on the physical parameters of the source stars.

\begin{table*}
\caption{Comparison of the $\chi^2_\mathrm{d.o.f}$ and BIC statistics. \label{tab:chi2_ET}}
\centering 
\tiny
\begin{tabular}{l|cc|lcc}
\hline\hline 
&\multicolumn{2}{c|}{w/o secondary peak} & \multicolumn{3}{c}{entire event}\\
Standard parametrization (Std) & $\chi^2_{\mathrm{d.o.f}}$ & BIC & ID$^c$& $\chi^2_{\mathrm{d.o.f}}$  & BIC \\
\hline 
1L1S + \PX~+ \XP~+  \PriorFlux$^a$& $674.45/611=1.104$ & $777.51$ & $-$ & $-$ & $-$ \\
1L2S + \PX~+ \XP~+ \PriorFlux$^a$ & $661.43/608=1.088$ & $783.81$ & $-$ & $-$ & $-$ \\
2L1S + \PX~+ \XP~+ \FS~+ \LD~ + \PriorFlux$^a$ & $-$ & $-$ & Std: 2L1S & $690.14/654=1.055$ & $820.40$ \\
2L2S + \PX~+ \XP~+ \FS~ + \LD ~+ \PriorFlux$^a$~+ \PriorSize$^b$& $-$ & $-$ & Std: 2L2S & $678.30/650=1.044$ & $838.29$ \\
\hline
Astrophysical source parametrization (ET) & & & & & \\
\hline 
1L1S + \PX~+ \XP~+ \FS~+ \LD   & $674.57/607=1.111$ & $803.39$ & $-$ & $-$ & $-$ \\
1L2S + \PX~+ \XP~+ \FS~+ \LD & $671.55/606=1.108$ & $806.81$ & $-$ & $-$ & $-$ \\
2L1S + \PX~+ \XP~+ \FS~ + \LD & $-$ & $-$ & ET: 2L1S & $695.58/651=1.068$ & $861.61$ \\
2L2S + \PX~+ \XP~+ \FS~+ \LD   & $-$ & $-$ & ET: 2L2S & $707.61/650=1.089$ & $883.50$ \\

\hline 
\end{tabular}\tablefoot{
\tablefoottext{a}{{non negative blending flux}}
\tablefoottext{b}{$f_{\mathrm{S1}}/f_{\mathrm{S2}}\approx(\rho_1/\rho_2)^2$}
\tablefoottext{c}{as in Figures \ref{fig:model_ET}, \ref{fig:model_zoom_ET}}
}
\end{table*}

\subsection{Results}
\subsubsection{Source system}
In the case of the 2L1S + \PX~+ \XP~ models, two solutions with the lowest $\chi^2$ values correspond to the same \XP~orbital solution
and differ only in the prominent inner/outer degeneracy. Solutions differ by $\Delta \chi^2 = 5.4$. The microlensing and physical parameters of the source resulting from the best-fitting model (ET: 2L1S) are listed in Table~\ref{tab:mulens_ET}.

\begin{figure}
    \centering
    \includegraphics[width=1\linewidth]{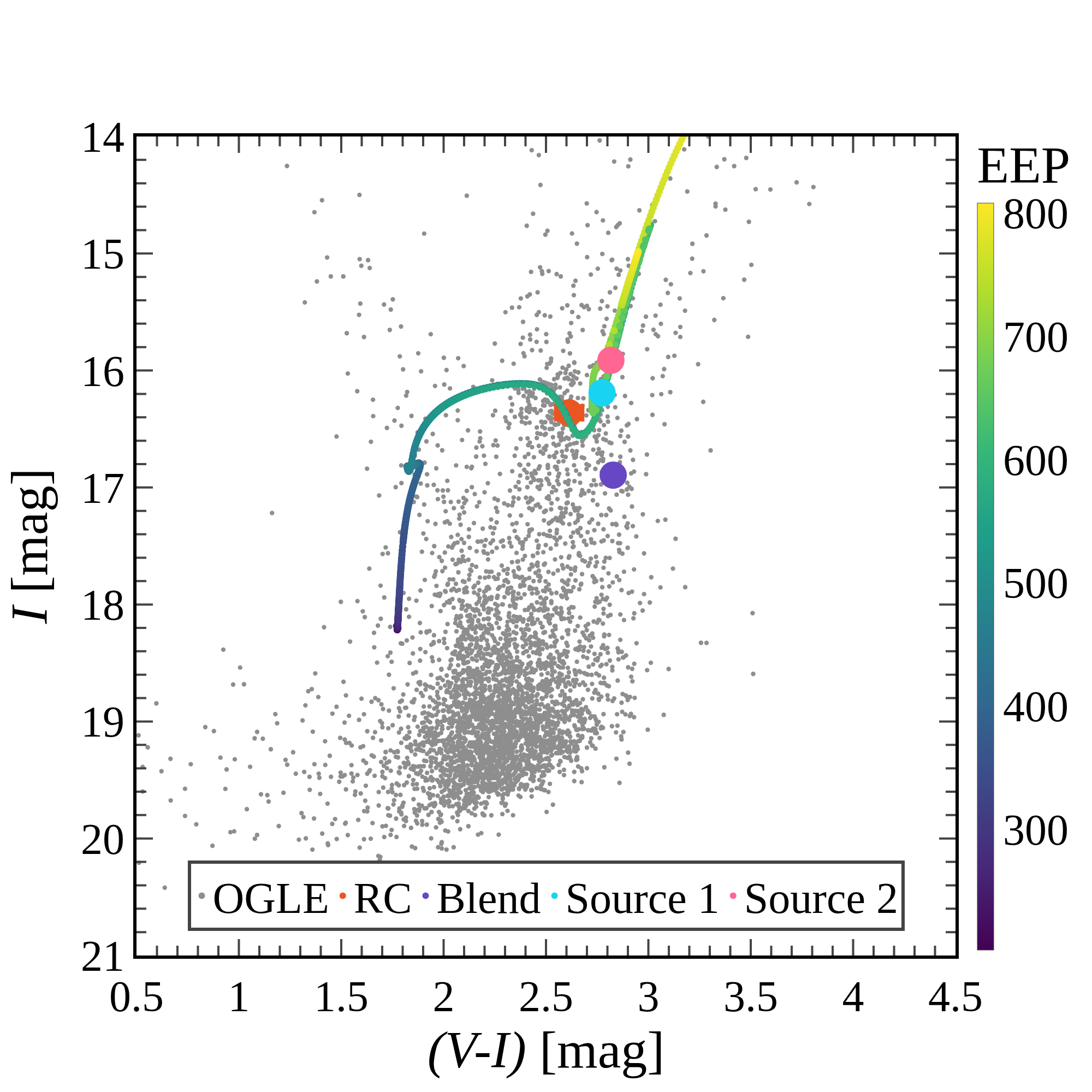}
    \caption{Color–magnitude diagram for stars from the OGLE-IV data within $3\arcmin$ of the microlensed star in the OGLE-2017-BLG-0114 event. The orange circle marks the position of the red clump (RC) centroid, while the pink and blue circles indicate the positions of the source stars, and the violet circle marks the blend. The evolutionary track of the source is shown as a line, with different colors representing successive evolutionary phases; the tracks for both sources overlap almost entirely. All values correspond to a binary-source model that includes the xallarap effect, fitted using astrophysical source parametrization (ET: 2L2S).}
    \label{fig:CMD_ET}
\end{figure}

Similarly, for the 2L2S + \PX~+ \XP~models, the three lowest-$\chi^2$ solutions correspond to the same \XP~orbital solution.  In terms of the physical parameters of the source systems, these three best solutions correspond to models where sources have nearly equal masses  ($q_{\mathrm{ source}} \approx 1$, with $M_{\rm init,S1} \approx 2.8 M_\odot$), both are red giant clump stars ($\mathrm{EEP_{S1}} \approx 660$) at a distance of $D_{\rm S} \approx 13$ kpc. Figure~\ref{fig:CMD_ET} presents the evolutionary tracks of the two sources for ET: 2L2S model, which has the lowest $\chi^2$ value, in comparison to the other stars within 3' from the event. The microlensing and physical parameters of the source are listed in Table~\ref{tab:mulens_ET}.

\subsubsection{Lens system}
The samplings of both the 1S and 2S models favor close configurations of the lens system, with $0.2886\pm0.0031$ and  $0.3242^{+0.0076}_{-0.0066}$  for the best-fitting 1S and 2S models, respectively. This finding is in agreement with the previous preliminary analysis by \cite{2021AcA....71....1P}.

We estimated the physical properties of the lens system following the procedure for standard parametrization models (Section 5.2.2), with two modifications: $\theta_{\rm E}$ was computed directly as the ratio of $\rho$ to the physical angular size of the source from MIST models (Equation \ref{equ:theta_size}), and the distribution $D_{\rm S}$ was taken directly from the MCMC sampling  rather than fixed at 8.54 kpc.

Mass estimates of the lens components depend on the adopted source system model. 
The best solution for the 2S models (ET: 2L2S) result in a notably lower value of $M_{\rm c}$. Nevertheless, the estimates of $M_{\rm c}$ agree within $1\sigma$ for the best-fitting 1S and 2S models. This is because, for the 1S models, the upper limit on $\theta_{\mathrm{E}}$ is poorly constrained, which in turn weakens the constraint on $M_{\rm c}$. Both 1S and 2S models indicate a cold super-Jupiter planet in the lens system.

Similar correlations occur for $M_{\rm h}$. The 2S models yield a host mass lower by about an order of magnitude. This results from lower estimates of the microlensing parallax in the case of the 2S models and is most likely caused by a correlation with the xallarap effect. A detailed characterization of the lens system, including all possible configurations and available data sets, will be presented in a forthcoming publication.

\begin{figure}
    \centering
    \includegraphics[width=0.99\linewidth]{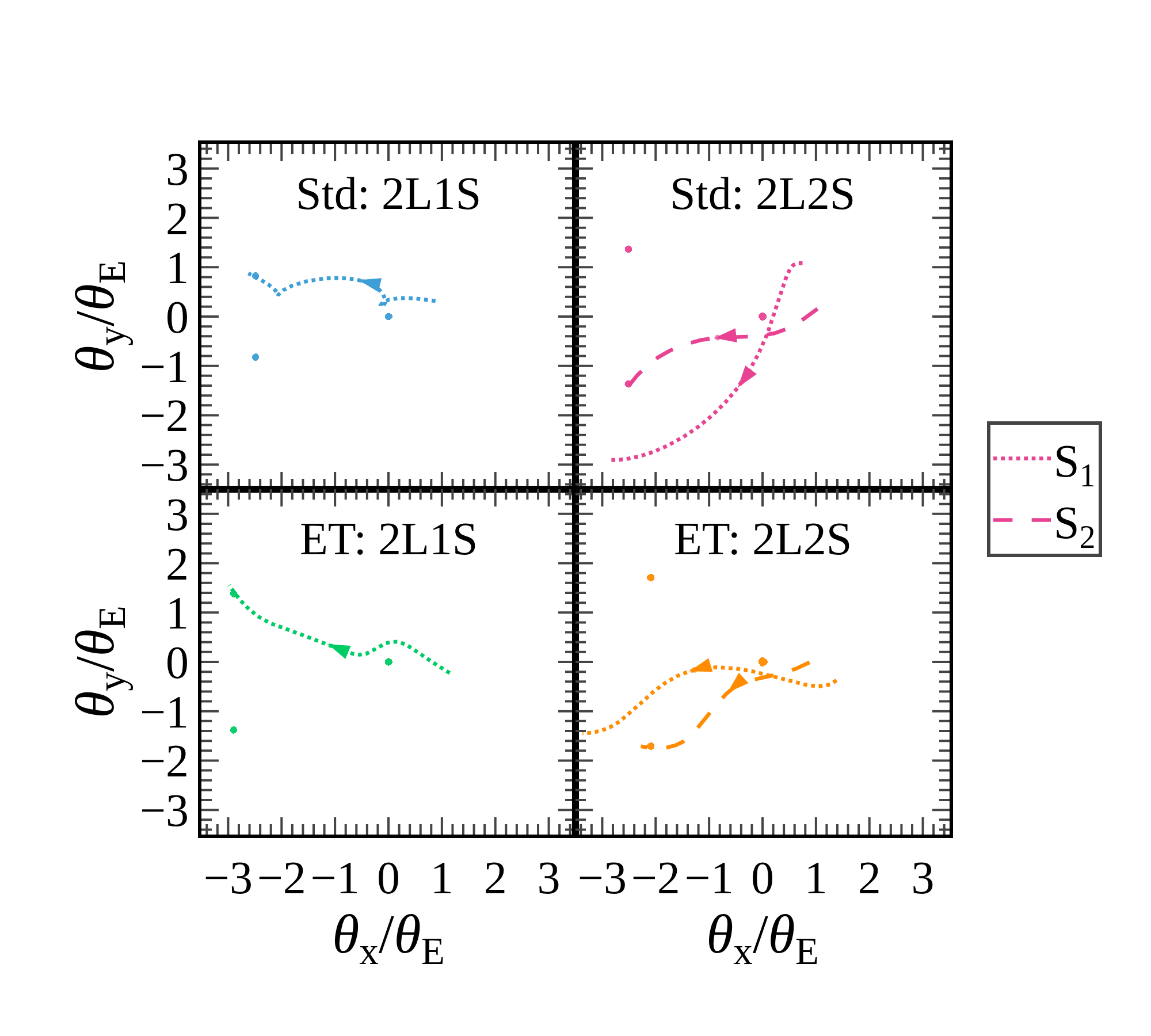}
    \caption{Caustics (small dots) and trajectories of sources (lines with arrows). All models include the xallarap effect. Note that all central caustics are at (0, 0) and planetary caustics have X coordinate between $-3$ and $-2$.}
    \label{fig:trj_ET_S}
\end{figure}
\begin{figure}
    \centering
    \includegraphics[width=0.99\linewidth]{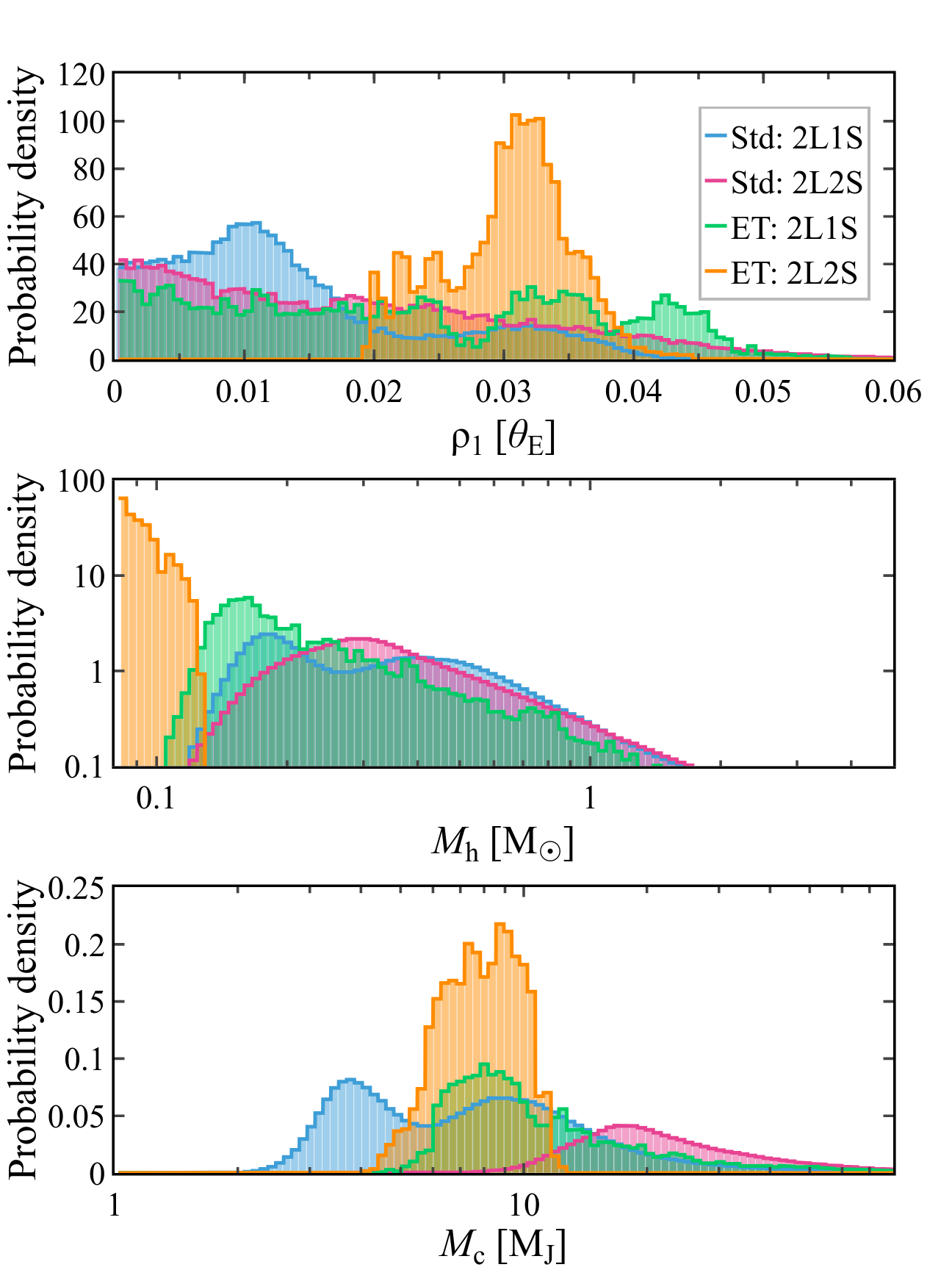}
\caption{Posterior distributions of the normalized source radius of primary source ($\rho_1$) from the MCMC samples and the corresponding distributions of the host ($M_{\mathrm{h}}$) and companion ($M_{\mathrm{c}}$) masses for OGLE-2017-BLG-0114. All models include the xallarap effect.}
    \label{fig:rho_ET_S}
\end{figure}
\section{Discussion on the method}
We have shown that incorporating a physical parametrization of the source system can be highly beneficial in the case of modeling complex microlensing events. Firstly, it allows us to impose additional constraints on the microlensing parameters, as we did in the case of 2S models (Equation~\ref{equ:prior_theta}). This, in turn, enabled us to improve the uncertainty of the $\theta_E$ measurements by an order of magnitude. Since stellar evolutionary models are better constrained in the presence of multi-band photometric observations, this approach should be even more beneficial in the modeling of events for which such time-series are available. Secondly, incorporating Galactic model constraints directly into the initial state of MCMC selection, or as priors in the fitting procedure, can offer advantages over other approaches in which physical parameters are inferred in a Bayesian manner based on the posterior of the chosen solution and Galactic model. Since the selection of the best solution can be somewhat arbitrary, it can affect the inferred parameters. We argue that in our approach, we gain better robustness, since our fitting does not rely solely on the $\chi^2$ statistic but also incorporates astrophysical constraints, and the astrophysical plausibility of degenerate solutions can be assessed at each step of the process.
Thirdly, thanks to the modern implementation of stellar evolutionary model interpolation in the {\tt brutus} code, incorporating a physical parametrization of the source is relatively computationally inexpensive. Sampling binary-source models with this approach requires, on average, roughly three times more computation time than the standard approach, but it can eliminate many candidate solutions already at the stage of generating initial conditions for the sampler, thereby reducing the overall computational cost.
Although this approach introduces additional parameters into the fitting procedure, it provides straightforward access to the constraints between them, which are contained in stellar evolutionary or Galactic models.

The presented method will become even more valuable in the era of upcoming space-based microlensing surveys. Missions of the Nancy Grace Roman Space Telescope \citep{2022BAAS...54e.146G} and Earth 2.0 \citep{2022arXiv220606693G} are expected not only to increase the number of detected microlensing events, but also to provide  the stability, precision, and continuity characteristic of space-based observations. As a result, higher-order effects, including xallarap and binary-source signatures, are likely to be detected more frequently and with greater significance. In particular, in combination with additional multi-band observations from programs such as the Legacy Survey of Space and Time \citep{2019ApJ...873..111I} conducted by the Vera C.\ Rubin Observatory, the method presented here can serve as a powerful tool for modeling and interpreting these events in an astrophysically meaningful way.

\section{Summary}
We present a novel method for microlensing modeling that simultaneously estimates the physical parameters of the source system using stellar evolutionary models. 

We implement this approach within a Markov Chain Monte Carlo framework and demonstrate its performance using the microlensing event OGLE-2017-BLG-0114 as a case study. We compare results obtained with the standard microlensing parametrization to those derived using the physical source parametrization, considering models that include binary lenses and source orbital motion (xallarap), with both single and binary source configurations.

We show that this approach enables the construction of fully astrophysically consistent models of both the lens and source systems. By enforcing physical constraints between xallarap parameters and the source properties, in the case of a binary source, we achieve an improvement in the estimates of $\theta_E$ by an order of magnitude.

Our results demonstrate that the adopted source system model can have a critical impact on the inferred properties of the lens system, including the mass of the host and its companion. This highlights the importance of incorporating physically motivated constraints directly into the modeling process, especially for complex microlensing events.

\begin{acknowledgements}
We thank the anonymous referee for their careful reading of the manuscript and their constructive comments, which helped improve this work. We are grateful to Sebastiano Calchi Novati for constructive comments and suggestions that significantly improved the clarity and interpretation of this work. OGLE Team acknowledges  former members of the team, for their contribution to the collection of the OGLE photometric data over the past years. The work of Mateusz Jan Mróz was partially supported by the Early Universe Scholarships funded by the Excellence Initiative – Research University program at the University of Warsaw. This research was funded in part by the National Science Center, Poland, grant Sonata Bis 2021/42/E/ST9/00038 to Radoslaw Poleski. 
\end{acknowledgements}
\bibliographystyle{aa}  
\bibliography{bib}
\FloatBarrier
\begin{appendix}
\section{Transformation between predicted location of planetary caustic cusps and microlensing parameters: $(\delta r, \delta \phi) \longrightarrow (s, \alpha)$}
To optimize the search for all degenerate solutions of the binary-lens model with planetary mass-ratio, we reparametrized the projected separation of the lenses, $s$, and the angle between the source trajectory and the binary axis, $\alpha$, using two parameters describing small, polar corrections to the predicted location of the planetary caustic cusp ($\delta r$, $\delta \phi$).
\par
\begin{figure*}
    \centering
    \includegraphics[width=0.99\linewidth]{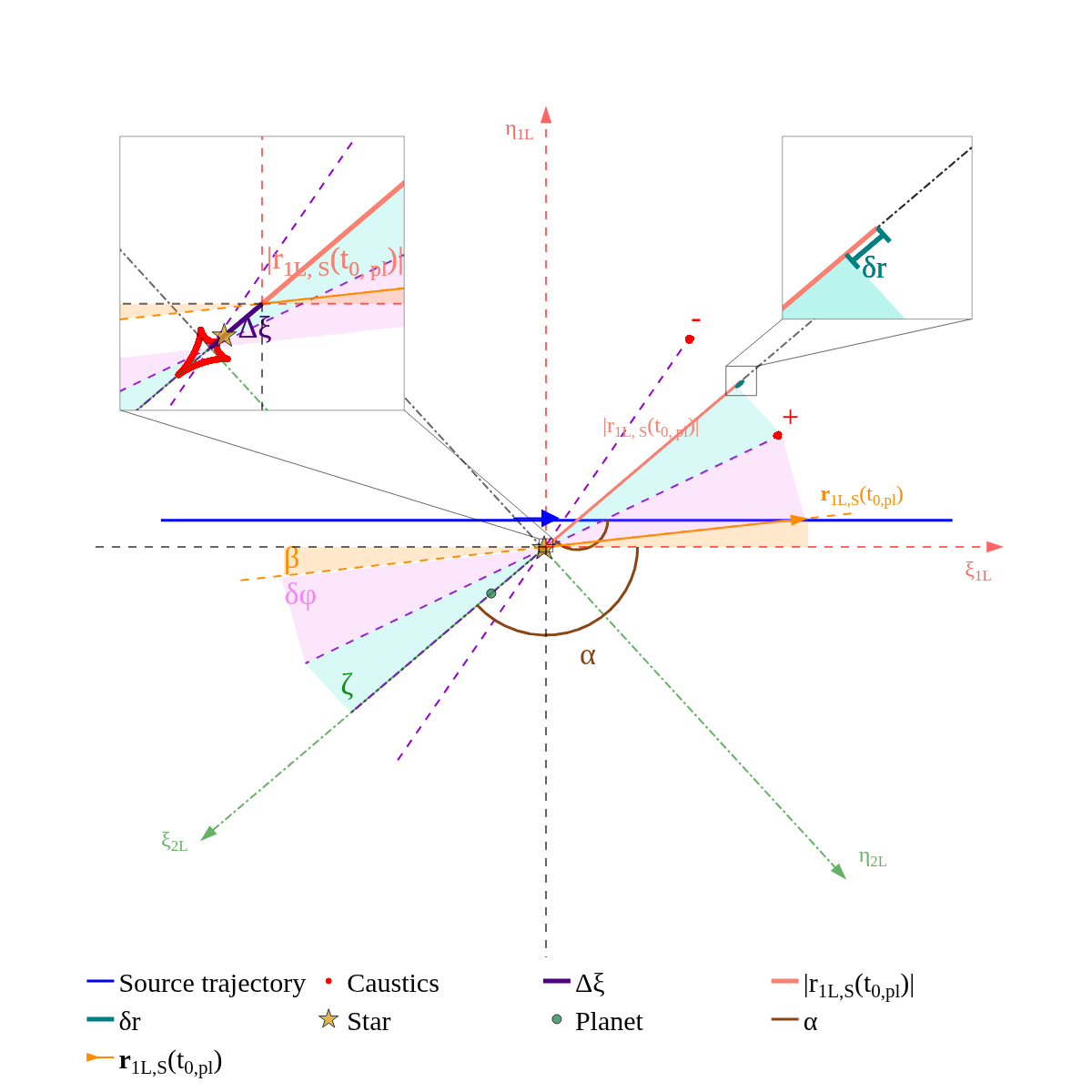}
    \caption{Observers view on the event geometry for close/inner topology. The lens star (yellow
star symbol) is near the origin. The planet is marked by green circle. The
source trajectory is indicated by a blue line which forms angle $\alpha$
(brown; its corresponding angle is also marked) with the lens axis. We
indicate the coordinate frame for both the single lens model (dashed
red lines $\xi_{1L}$ and $\eta_{1L}$; parallel to the source trajectory)
and the binary-lens model (dashed-dotted green lines $\xi_{2L}$ and
$\eta_{2L}$; parallel to the lens axis). For both coordinate frames, the negative directions of all four axes are shown in gray.  Source position is indicated by the
orange vector $\mathbf{r}_{\mathrm{1L,S}}(t_{\mathrm{0,pl}})$. This vector forms
angle $\beta$ with $\xi_{1L}$. Caustics are marked with red color: central one is seen in the left zoom-in and the two planetary caustics are on the right-hand side of the plot. The upper cusp of the planetary caustic is marked with a ``$+$'' sign, and the bottom cusp with a ``$-$'' sign. Vectors pointing to their locations are indicated by violet dashed lines. The vector corresponding to the upper caustic forms an angle $\zeta$ with $\xi_{\mathrm{2L}}$ and an angle $\delta\phi$ with the source trajectory $\mathbf{r}_{\mathrm{1L,S}}(t_{0,\mathrm{pl}})$. The length of the vector $\mathbf{r}_{\mathrm{1L,S}}(t_{0,\mathrm{pl}})$ projected onto the $\xi_{\mathrm{2L}}$ axis is marked with a solid salmon line. The approximate shift between the origins of the coordinate frames, $\Delta\xi$, is marked with a solid dark violet line. The difference between the actual location of the planetary caustic axis and the sum of the lengths $\Delta\xi$ and $|\mathbf{r}_{\mathrm{1L,S}}(t{_0,\mathrm{pl}})|$, denoted by $\delta r$, is marked with a solid green line.
}
\label{fig:angels}
\end{figure*}

For this purpose, we combined information about the position of the source (or sources, in the case of a binary-source model) derived from the single-lens model at any given moment with the light-curve constraint indicating that the source approached the planetary caustic cusp at $t_{0, \mathrm{pl}}$. We further used the known relations between ($s$, $q$) and the location of the upper/lower (off-binary-axis) planetary caustic cusps \citep{2006ApJ...638.1080H}.
We introduced two parameters instead of ($s$, $q$). 
First, $\delta r$ is the distance between the source position at $t_{0,\mathrm{pl}}$, rotated onto the binary axis (by the angle $\alpha$), and the center of the planetary caustic (the $\xi$-axis in \citealt{2006ApJ...638.1080H}; not to be confused with the $\xi$ with a subscript used elsewhere in this work to describe the xallarap orbit parameters.) -- see Fig.~\ref{fig:angels}. 
Second, $\delta \phi$ is the angle between the source position at $t_{0,\mathrm{pl}}$ and the position of the upper or lower planetary caustic cusp in the single-lens coordinate system.
The first step in the sampling was to calculate the trajectory of the source in the single-lens (1L) coordinate system, i.e., to compute the trajectory assuming a model without the binary-lens parameters ($s$, $q$, and $\alpha$). We denote the source position in this coordinate system as $\mathbf{r}_{\mathrm{1L, S}}$. The shift between the magnification center estimated from the single-lens model and the actual center of mass of the binary system can be approximated by (A.18 in \citealp{2011ApJ...738...87S}):
\begin{equation}
\Delta\xi  \approx
\frac{q}{1+q} \left(\frac{1}{s(t_{0,\mathrm{2L}})} -s(t_{0,\mathrm{2L}})\right). 
\end{equation}
Here, $\Delta\xi$ follows the axis notation of \cite{2006ApJ...638.1080H}. 
Assuming the distance between the primary lens and the center of the planetary caustic is $|s - 1/s|$, then:
\begin{equation}\begin{cases}
    \left(s - \frac{1}{s}\right) + \Delta\xi =
\left|\mathbf{r}_{\mathrm{1L, S}}(t_{\mathrm{0, pl}}) \right| + \delta r  & \text{for } s > 1, \\ 
    \left(s - \frac{1}{s}\right) - \Delta\xi =
-  \left(\left|\mathbf{r}_{\mathrm{1L, S}}(t_{\mathrm{0, pl}}) \right| + \delta r \right)  & \text{for } s < 1.
\end{cases}
\end{equation}
From this relation, $s$ can be expressed as:
\begin{equation}
s =d +\sqrt{d^2 + 1}, 
\end{equation}
where
\begin{equation}
d =
\begin{cases}
\frac{1}{2} \left(1+q\right)\left( \left|\mathbf{r}_{\mathrm{1L, S}}(t_{\mathrm{0, pl}}) \right| + \delta r \right) & \text{for } s > 1, \\
- \frac{1}{2} \frac{1+q}{1+2q}\left(\left|\mathbf{r}_{\mathrm{1L, S}}(t_{\mathrm{0, pl}}) \right| + \delta r \right) & \text{for } s < 1.\\
\end{cases}
\end{equation}

Once $s$ is determined from the above relations, the next step is to recover the angle $\alpha$. For this purpose, we define the following auxiliary angles (see Fig. \ref{fig:angels}):
$\beta$ -- the angle between the position vector $\mathbf{r}_{\mathrm{1L, S}}(t_{\mathrm{0, pl}})$ and the $x$-axis of the 1L coordinate system (i.e., the axis parallel to the relative proper motion). It is related to the components of $\mathbf{r}_{\mathrm{1L, S}}(t_{\mathrm{0, pl}})$ by
\begin{equation}
\tan \beta = \frac{y_{\mathrm{1L, S}}(t_{\mathrm{0, pl}})}{x_{\mathrm{1L, S}}(t_{\mathrm{0, pl}})} ,
\end{equation}

$\zeta$ -- the angle between the position of the upper or lower planetary caustic cusp and the $x$-axis of the 2L coordinate system (i.e., the binary axis), expressed as
\begin{equation}
\begin{cases}
 \tan \zeta \approx\frac{\pm \frac{2 q^{1/2}}{s\sqrt{s^2+1}}}{(s - 1/s) + \Delta\xi}    & \text{for } s > 1, \\
  \tan \zeta \approx\frac{\pm \frac{2 q^{1/2}}{s\sqrt{s^2+1}}}{(s - 1/s) - \Delta\xi}    & \text{for } s < 1,
\end{cases}
\end{equation}
where the $\pm$ sign corresponds to the cusp located above ($+$) or below ($-$) the binary axis.
Finally, angle $\alpha$ can be obtained as:
\begin{equation}
\alpha = 180^\circ - \beta - \zeta - \delta\phi.
\end{equation}

\FloatBarrier 
\onecolumn
\section{Model parameters for OGLE-2017-BLG-0114}
\begin{table*}[h!]
\centering
\tiny
\caption{Microlensing xallarap model parameters for OGLE-2017-BLG-0114.} \label{tab:mulens_ET}
\begin{tabular}{l|ll|ll}
\toprule
 Parametrization & \multicolumn{2}{c|}{Standard (Std)}  & \multicolumn{2}{c}{Astrophysical source (ET) }  \\
 \midrule
Model & 2L1S + \PX  + \XP & 2L2S + \PX  + \XP & 2L1S + \PX  + \XP& 2L2S + \PX  + \XP\\
Degeneracy & $u_{{0}}{{-}}$, 1, close, \sP , inner & $u_{{0}}{{+}}$, 5, close, \sS , outer & $u_{{0}}{{-}}$, 0, close, \sP , inner & $u_{{0}}{{+}}$, 11, close, \sS , outer \\
$t_0$ & $2457899.3^{+2.7}_{-3.4}$ & $2457947.7^{+1.3}_{-1.1}$ & $2457901.49\pm0.15$ & $2457931.76\pm0.34$ \\
$u_0$ & $-0.144\pm0.023$ & $0.156\pm0.013$ & $-0.0582\pm0.0081$ & $0.214^{+0.012}_{-0.013}$ \\
$t_{\rm E}$ & $173\pm16$ & $112.8^{+7.1}_{-5.0}$ & $92.54\pm0.28$ & $95.0^{+2.9}_{-3.9}$ \\
$\rho_1$ & $0.0109^{+0.0140}_{-0.0070}$ & $0.016^{+0.017}_{-0.012}$ & $0.022^{+0.018}_{-0.016}$ & $0.0306^{+0.0039}_{-0.0065}$ \\
$\rho_2$ & $\text{---}$ & $0.0332^{+0.0095}_{-0.0159}$ & $\text{---}$ & $0.0348^{+0.0077}_ {-0.0090}$ \\
$\pi_{\mathrm{E,N}}^{a}$ & $0.167^{+0.023}_{-0.026}$ & $-0.091^{+0.031}_{-0.025}$ & $0.197^{+0.026}_{-0.035}$ & $0.284\pm0.020$ \\
$\pi_{\mathrm{E,E}}^{a}$ & $0.127\pm0.023$ & $0.041\pm0.012$ & $-0.052\pm0.013$ & $-0.101^{+0.016}_{-0.019}$ \\
$s$ & $0.337\pm0.016$ & $0.3138^{+0.0124}_{-0.0080}$ & $0.2886\pm0.0031$ & $0.3242^{+0.0076}_{-0.0066}$ \\
$q$ & $0.0219^{+0.0054}_{-0.0039}$ & $0.060^{+0.015}_{-0.011}$ & $0.0506\pm0.0093$ & $0.100^{+0.023}_{-0.019}$ \\
$\alpha$ & $330.4\pm2.6$ & $42.5^{+3.9}_{-4.6}$ & $312.2\pm3.4$ & $352.1^{+2.8}_{-3.2}$ \\
$\xi_{\mathrm{P}}$ & $221.5\pm4.2$ & $624^{+55}_{-39}$ & $658\pm28$ & $700\pm52$ \\
$\xi_{\mathrm{a}}$ & $0.200^{+0.033}_{-0.025}$ & $1.03^{+0.13}_{-0.11}$ & $0.807^{+0.024}_{-0.028}$ & $0.437^{+0.046}_{-0.051}$ \\
$\xi_{\mathrm{\Omega}}$ & $250.0^{+15.8}_{-7.6}$ & $54.6^{+5.7}_{-4.6}$ & $110.0\pm1.0$ & $158.9\pm5.0$ \\
$\xi_{\mathrm{i}}$ & $30.5^{+7.0}_{-8.1}$ & $106.7\pm2.9$ & $84.14\pm0.32$ & $71.3^{+3.0}_{-5.0}$ \\
$\xi_{\mathrm{u}}$ & $236.1^{+7.4}_{-17.0}$ & $257.3^{+2.0}_{-2.7}$ & $274.4\pm2.4$ & $89.5\pm3.4$ \\
$\xi_{\mathrm{e}}$ & $0.615^{+0.015}_{-0.018}$ & $0.173\pm0.099$ & $0.7731\pm0.0089$ & $0.345\pm0.032$ \\
$\xi_{\mathrm{\omega}}$ & $207.3^{+8.7}_{-16.1}$ & $169^{+12}_{-16}$ & $266.3\pm1.2$ & $33.0\pm8.1$ \\
$q_{\mathrm{source}}$ & $\text{---}$ & $1.86^{+0.39}_{-0.26}$ & $\text{---}$ & $0.9965^{+0.0010}_{-0.0012}$ \\
$M_{\mathrm{init,S}1}~\mathrm{[M_{\odot}]}$ & $\text{---}$ & $\text{---}$ & $2.58^{+0.53}_{-0.40}$ & $2.879\pm0.073$ \\
$\text{EEP}_{\mathrm{S}1}$ & $\text{---}$ & $\text{---}$ & $618^{+11}_{-19}$ & $662.1^{+3.1}_{-5.8}$ \\
$[\mathrm{Fe/H}]_{\mathrm{S}1}$ & $\text{---}$ & $\text{---}$ & $-0.60\pm0.51$ & $-0.158^{+0.108}_{-0.088}$ \\
$D_{\rm S}~\mathrm{[kpc]}$ & $\text{---}$ & $\text{---}$ & $11.2^{+3.9}_{-2.4}$ & $14.0^{+1.9}_{-1.3}$ \\
$A_V~\mathrm{[mag]}$ & $\text{---}$ & $\text{---}$ & $3.72^{+0.19}_{-0.33}$ & $3.700^{+0.059}_{-0.084}$ \\
$f_{\mathrm{S}1,I-\mathrm{OGLE}}$ & $398^{+77}_{-64}$ & $179\pm15$ & $630.0^{+5.0}_{-5.8}$ & $193\pm13$ \\
$f_{\mathrm{S}2,I-\mathrm{OGLE}}$ & $\text{---}$ & $400^{+35}_{-52}$ & $\text{---}$ & $327\pm35$ \\
$f_{\mathrm{B},I-\mathrm{OGLE}}$ & $230^{+63}_{-77}$ & $44^{+65}_{-36}$ & $-1.8^{+5.9}_{-4.9}$ & $110^{+31}_{-35}$ \\
$\chi^{2}~/~\mathrm{d.o.f.}$ & 690.1 / 654 & 678.3 / 650 & 715.9 / 651 & 707.6 / 650 \\
$\ln{\Prior}$ & 0.00 & -1.84 & -8.19 & -9.79 \\
BIC & 820.40 & 838.29 & 882.12 & 883.50 \\
\bottomrule
\end{tabular}
\tablefoot{
\tablefoottext{a}{For the reference time $t_{\mathrm{0,par}}=2457933.5$~HJD}
}
\end{table*}
\begin{table*}[h!]
\centering
\tiny
\caption{Physical parameters of xallarap models for OGLE-2017-BLG-0114.} \label{tab:parms_lens_ET} 
\begin{tabular}{l|ll|ll}
\toprule
 Parametrization & \multicolumn{2}{c|}{Standard (Std)}  & \multicolumn{2}{c}{Astrophysical source (ET) }  \\
 \midrule
Model & 2L1S + \PX  + \XP & 2L2S + \PX  + \XP & 2L1S + \PX  + \XP& 2L2S + \PX  + \XP\\
Degeneracy & $u_{{0}}{{-}}$, 1, close, \sP , inner & $u_{{0}}{{+}}$, 5, close, \sS , outer & $u_{{0}}{{-}}$, 0, close, \sP , inner & $u_{{0}}{{+}}$, 11, close, \sS , outer \\
$\theta_E~\mathrm{[mas]}$ & $0.86^{+1.56}_{-0.49}$ & $0.36^{+0.71}_{-0.14}$ & $0.50^{+1.42}_{-0.22}$ & $0.184^{+0.040}_{-0.019}$ \\
$\mu_{\rm{rel}}~\mathrm{[mas/yr]}$ & $1.8^{+3.3}_{-1.0}$ & $1.14^{+2.24}_{-0.45}$ & $1.91^{+5.46}_{-0.84}$ & $0.689^{+0.152}_{-0.068}$ \\
$M_{\rm{h}}~\mathrm{[M_{\odot}]}$ & $0.50^{+0.90}_{-0.28}$ & $0.47^{+1.02}_{-0.21}$ & $0.31^{+0.92}_{-0.15}$ & $0.0743^{+0.0172}_{-0.0090}$ \\
$M_{\rm{c}}~\mathrm{[M_{J}]}$ & $11.5^{+21.7}_{-6.1}$ & $30^{+59}_{-14}$ & $16.3^{+50.9}_{-8.3}$ & $8.2\pm1.8$ \\
$a_{\mathrm{L},\perp}~\mathrm{[au]}$ & $0.97\pm0.32$ & $0.74^{+0.81}_{-0.24}$ & $0.76^{+0.46}_{-0.23}$ & $0.478^{+0.050}_{-0.040}$ \\
$D_{\rm{L}}~\mathrm{[kpc]}$ & $3.4\pm1.8$ & $6.50^{+0.70}_{-1.91}$ & $5.1^{+1.9}_{-3.0}$ & $7.81^{+0.97}_{-0.87}$ \\
$a_{\mathrm{S1}}~\mathrm{[au]}$ & $1.46^{+2.86}_{-0.81}$ & $3.2^{+6.4}_{-1.3}$ & $4.4^{+13.9}_{-2.0}$ & $1.163\pm0.057$ \\
$K_{\rm{RV,S1}}~\mathrm{[km/s]}$ & $44^{+96}_{-25}$ & $54^{+106}_{-22}$ & $115^{+364}_{-52}$ & $18.0^{+2.0}_{-1.6}$ \\
$I_{\rm{S1}}~\mathrm{[mag]}$ & $15.31^{+0.19}_{-0.38}$ & $16.276^{+0.091}_{-0.182}$ & $14.9924^{+0.0084}_{-0.0186}$ & $16.219^{+0.068}_{-0.146}$ \\
$I_{\rm{S2}}~\mathrm{[mag]}$ & $\text{---}$ & $15.404^{+0.090}_{-0.240}$ & $\text{---}$ & $15.61^{+0.11}_{-0.23}$ \\
$\chi^{2}~/~\mathrm{d.o.f.}$ & 690.1 / 654 & 678.3 / 650 & 715.9 / 651 & 707.6 / 650 \\
$\ln{(\Prior)}$ & 0.00 & -1.84 & -8.19 & -9.79 \\
BIC & 820.40 & 838.29 & 882.12 & 883.50 \\
\bottomrule
\end{tabular}
\end{table*}

\end{appendix}
\end{document}